
\input harvmac
\def\title#1#2#3#4{\Title{ESENAT-#1}{#2}\vskip -0.3in
\centerline{{\titlefont#3}}\vskip 0.20in
\centerline{{\titlefont#4}}\vskip 0.4in
}
\def\jsp{\centerline{Jae-Suk Park\footnote{$^\dagger$}
{Bitnet: pjesenat@krysucc1}}\bigskip\centerline{{\it ESENAT}}
\centerline{{\it Research Institute for Theoretical Physics}}
\centerline{{\it 70-30 Changcheon-Dong, Seodamun-Gu}}
\centerline{{\it Seoul 120-180, Korea}}
\bigskip\bigskip\bigskip\bigskip}
\def\abs#1#2{\centerline{{\bf ABSTRACT}}\vskip 0.3in
{#2}\Date{#1}}

%
\def\a{\alpha}    \def\b{\beta}       \def\c{\chi}       \def\d{\delta}
\def\D{\Delta}     \def\f{\phi}       \def\F{\Phi}
\def\g{\gamma}               
                    
             \def\p{\psi}
\def\P{\Psi}            \def\S{\Sigma}

%

\def\CA{{\cal A}}   
 \def\CF{{\cal F}}  \def\CG{{\cal G}}

   \def\CU{{\cal U}}
\def\CW{{\cal W}} \def\CM{{\cal M}}
%
\def\rd{\partial}

\def\darr#1{\raise1.5ex\hbox{$\leftrightarrow$}\mkern-16.5mu #1}

\def\ha{{\textstyle{1\over2}}}

\def\Fr#1#2{{#1\over#2}}
\def\tr{\hbox{Tr}\,}

%
\def\cmp#1#2#3{, Comm.\ Math.\ Phys.\ {{\bf #1}} {(#2)} {#3}}

\def\np#1#2#3{, Nucl.\ Phys.\ {{\bf #1}} {(#2)} {#3}}

\def\ijmp#1#2#3{, Int.\ J.\ Mod.\ Phys.\ {{\bf #1}} {(#2)} {#3}}

\def\jdg#1#2#3{, J.\ Diff.\ Geo.\ {{\bf #1}} {(#2)} {#3}}
\def\pnas#1#2#3{, Proc.\ Nat.\ Acad.\ Sci.\ USA.\ {{\bf #1}} {(#2)} {#3}}

\def\zp#1#2#3{, Z.\ Phys.\ {{\bf #1}} {(#2)} {#3}}

\def\ptrsls#1#2#3{, Philos.\ Trans.\  Roy.\ Soc.\ London Ser.\
{{\bf #1}} {(#2)} {#3}}

%
\def\dt{\d_{\!{}_{T}}}
\def\dw{\d_{\!{}_{W}}}
\def\db{\d_{\!{}_{B\!R\!S}}}
\def\tb{\tilde{\d}_{\!{}_{B\!R\!S}}}
\def\hCF{\widehat{\CF}}

\def\tCW{\widetilde{\CW}}
\def\tW{\widetilde{W}}
\def\uW{\underline{W}}
\def\uCW{\underline{\CW}}

\def\dA{d_{\CA}}

\def\hv{\hat{v}}


\lref\tg{
E.\ Witten\np{B340}{1990}{523}\semi
M.\ Kontsevich\cmp{147}{1992}{1}
}
\lref\tym{
E.\ Witten\cmp{141}{1991}{153}\semi
Two dimensional gauge theories revisited, IASSNS-HEP-92/15
}
\lref\W{
E.\ Witten\cmp{117}{1988}{353}}
\lref\WW{
E.\ Witten, Introduction to cohomological field theories,
{\it in\/} Proc.\ Trieste Conference on
Topological methods in quantum field theories (ICTP, Trieste, June
1990), ed.\ W.\ Nahm et.\ al., (World Scientific, Singapore, 1991)
}
\lref\D{
S.K.\ Donaldson\jdg{18}{1983}{269}; {\bf 26} (1987) 397;
Topology {\bf 29} (1990) 257}
\lref\Horne{
J.H.\ Horne\np{B318}{1989}{22}}
\lref\M{
R.\ Myers\ijmp{A5}{1990}{1369}}
\lref\Kanno{
H.\ Kanno\zp{C43}{1989}{477}}
\lref\AS{
M.F.\ Atiyah and I.M.\ Singer\pnas{81}{1984}{2597}
}
\lref\BRS{
C.\ Becchi, A.\ Rouet and R.\ Stora\cmp{42}{1975}{127}\semi
I.V.\ Tyutin, Lebedev Inst.\ preprint, (1975)
\semi G.\ Curci and R.\ Ferrari.\ Nuovo Cimento,{\bf A32} (1976) 151;
A35 (1976) 273}
\lref\Stora{
R.\ Stora, Algebraic structure and topological origin of
anomalies, {\it in\/} Progress in gauge field theory, ed.\ G.\
't Hooft et al.\ (Plenum, N.Y., 1984)}
\lref\ZuminoA{
B.\ Zumino, Chiral anomalies and differential geometry, {\it
in\/} Relativity, group and topology II, eds.\ B.S.\ deWitt et.\
al., Les Houches 1983, (North-Holland, Amsterdam, 1984)}
\lref\ZuminoC{
B.\ Zumino, Anomalies, cocycles and Schwinger terms,
{\it in\/} Symposium on anomalies, geometry and topology,
March 28-30, 1985,
eds.\ W.A.\ Bardeen and A.R.\ White (World Scientific, Singapore)}
\lref\BZ{
W.A.\ Bardeen and B.\ Zumino\np{B244}{1984}{421}}
\lref\Gri{V.\ Gribov\np{B139}{1978}{1} }
\lref\S{I.M.\ Singer\cmp{60}{1978}{7} }
\lref\AHS{M.F.\ Atiyah, N.J.\ Hitchin and
I.M.\ Singer\ptrsls{A308}{1982}{524}
}
\lref\Park{J.S.\ Park, Algebraic structures in topological Yang-Mills
theory, ESENAT-92-04}
\title{92-08}{Zero-Modes, Covariant Anomaly Counterparts}{
and Reducible Connections
}{
in Topological Yang-Mills Theory}

\jsp

\abs{October, 1992}{
We introduce the covariant forms for the non-Abelian anomaly counterparts
in topological Yang-Mills theory, which satisfies the topological descent
equation modulo terms that vanish at the space of BRST fixed points.
We use the covariant anomalies as a new set of observables, which can
absorb both  $\dw$ and $\db$ ghost number violations of zero-modes.
Then, we study some problems due to the zero-modes originated from the
reducible connections.
}


One of the most important conceptual revolutions of topological field
theories (TFT's) is the unbroken phase of the general covariance or that
of string theory is realized.
It has been shown that the cohomological TFT's in two dimensions are
equivalent to the physical ones\tg\tym.
Though it is tempting to believe that TFT's have opened new horizon of
unified theory of everything, almost nothing is known how to induce the
broken phases in the realistic dimensions.

Topological Yang-Mills theory (TYM)\W, which is the first example of
TFT's in the cohomological nature, has another open problem that
there is no known consistent mathematical or physical method to define
theory in the presence of reducible connections\D\WW.
Due to a BRST-like fermionic symmetry,
the entire path integral of a TFT localized to the locus
of the BRST fixed points (exact semi-classical limit), which is a certain
moduli space in general\W.
In TYM, the fermionic symmetry is generated by the Witten's $\dw$
operator, and the basic $\dw$ algebra is given by
\eqn\bach{
\dw\CA = \P,\quad\dw\P= -\dA\F,\quad\dw\F=0.
}
Using the $\dw$ symmetry one restrict the configuration space
of the theory to the space of instanton, which leads
to another $\dw$ algebra
\eqn\hendel{
\dw\c = \CF^+,
}
where the superscript $+$ denotes the projection to the self-dual parts
of the curvature $\CF$. The resulting $\dw$ fixed action has the usual
gauge symmetry, which can be fixed following the conventional BRS gauge
fixing method\Horne\M.
Then we should consider  $\db$ algebra\M
\eqn\guno{
\db\CA = -\dA v,\quad \db\bar{v} = \hbox{gauge fixing condition}
}
where $v,\bar{v}$  is the Faddev-Popov ghost and the anti-ghost
respectively.

Then, the space of the $\dw$ and $\db$ fixed points
$\dw\c=0,\;\dw\P=0,\;\db\CA=0,\;\db\bar{v}=0$
is precisely the moduli space of instanton $\CM$ with the space of $\F$
zero modes and that of $v$ zero-modes.
Thus, if there are no $\F$ and $v$ zero-modes, the path integral exactly
reduce to the moduli space of instanton. In this letter, we discuss
the problem of the $v,\F$ zero-modes originated
from the reducible connections.

Another important property of TYM is that the cohomological structures of
the instanton moduli space $\CM$ - which can be summarized by
the instanton complex\AHS
\eqn\mozart{0 \to A^{0}(ad\,
P)\mathrel{\mathop\to^{d_{\CA} }} A^{1}(ad\,
P)\mathrel{\mathop\to^{d^{+}_{\CA}}} A^{2}_{+}(ad\, P)
\to 0,
}
where $ad\, P = P \times_{ad}\,{\bf g}$ denotes the adjoint bundle
associated with a principle G-bundle $P$ over a compact oriented Riemann
4-manifold $M$, $A^p(F)$ denotes the space of $F$-valued p-forms and $+$
denotes projection to self-dual part,
are realized by the zero-modes of fields
$((\F,\bar\F, \eta),\P,\c)$ with the $\dw$ ghost number
$((2,-2,-1), 1 ,-1)$, whose
zero-modes are the solutions to the following equations
\eqn\zeromode{
\dA\F =\dA\bar\F=\dA\eta =0,\quad
\dA^*\P=\dA^{+}\P=0,\quad
\dA^{+*}\c = 0.
}
The number of non-trivial solutions for
$((\F,\bar\F,\eta),\P,\c)$ precisely correspond
to $(h^0, h^1, h^2)$,
where $h^{i} = \hbox{dim } H^{i}$ denotes the dimension ($i$-th
Betti number) of the cohomology groups of the instanton complex \mozart.
It follows that the net $\dw$ ghost number
violation of zero modes equals to the formal dimension of $\CM$
given by the index $s = h^1-h^0-h^2$ of the
instanton complex.

In particular, the
$0$th cohomology $H^0$ is non-trivial ($\F$ zero-modes) for the
reducible connection, and it is a source of singularities in the moduli
space $\CM$. And, the
elements of $1${\it th\/} cohomology group ($\P$ zero-modes)  can be
identified to the tangent vectors of $\CM$.
Thus, the formal dimension is the actual dimension when
$h^0 = h^2 = 0$. Then the moduli space is a smooth manifold,
and the Donaldson's invariants are well-defined.

Thus, the non-zero dimension of the instanton moduli space implies the
$\dw$ ghost number anomaly and appropriate set of observables
should be inserted to compensate it. Such set of observables
has been introduced by Witten\W\ and interpreted geometrically
based on the universal bundle\AS\ by Kanno\Kanno. From an obvious
candidate $\tCW_{0}{}^{0,4}=\ha\tr(\F^2)$, we can find the topological
descent equation after some iterations
\eqn\fuga{\eqalign{
0 &= \dw\tCW_{0}{}^{0,4},\cr
d\tCW_{0}{}^{0,4}&= \dw\tCW_{1}{}^{0,3},\cr
d\tCW_{1}{}^{0,3}&= \dw\tCW_{2}{}^{0,2},\cr
d\tCW_{2}{}^{0,2}&= \dw\tCW_{3}{}^{0,1},\cr
d\tCW_{3}{}^{0,1}&= \dw\tCW_{4}{}^{0,0},\cr
d\tCW_{4}{}^{0,0}&= 0,\cr
}}
where $\tCW_{k}{}^{0,4-k}$ denote $k$ form with $\db$ and $\dw$ ghost
number $0$ and $4-k$ respectively. Integrating $i$-th relation over
a $i-1$ dimensional cohomology cycle $\g_{{}_{i-1}}$, we can see
that
\eqn\rondo{
\dw \int_{\g_{{}_k}} \tCW_{k}{}^{0,4-k} \equiv \dw \tW^{0,4-k} =0
}
$\tW^{0,4-k}$ is $\dw$ closed. One can also easily see that
the Witten's observable is non-trivial and its $\dw$ cohomology
class depends only on the homology class of $\g_{{}_k}$.
It is also well known that $\tCW_{k}^{0,4-k}$ is a component
of the characteristic class $\ha\tr\hCF^2 =\ha \tr(\CF +\P +\F)^2$.
We can also start from a higher characteristic class
$\tCW_{0}{}^{0,2n} = c_n \tr \F^n$ and obtain $\tCW_{k}{}^{0,2n-k}$
after repeating the iteration \fuga. And, after integrating over
a $k_i$ dimensional  homology cycle $\g_{{}_{k}}$,
we get $\tW^{0,2n-k}$, which is an element of $2n-k$ cohomology
class on the orbit space $H^{2n-k}(\CU/\CG)$.

Note that the zero-modes of the Faddev-Popov ghost $v$, which are the
non-trivial solutions of $\dA v=0$, as well as those of $\F$ can be
originated from the reducible connections. Because, in the Horne's
approach\Horne\M\Park, both $\dw$ and $\db$ ghost numbers should be
preserved separately and the completely fixed action of TYM has
both $\dw$ and $\db$ ghost number zero,
we should also insert appropriate set of observables to absorb the net
violation of the $\db$ ghost number.
Such an observable should be $\dw$ as well as $\db$ closed and non-trivial
in the sense of global topology. If there is no such an observable,
not only the topological interpretation become impossible but also
the correlation function itself can not be
well defined.

A way out of this problem was proposed by the author using the
non-Abelian (consistent) anomaly counterparts in TYM, which can be obtained
from one higher rank characteristic class $\tr\hCF^3$ using the
extended descent equation\Park. If one insert the consistent anomaly, which
is only $\db$ closed but has $\db$ ghost number $1$, in addition to the
Witten observables to absorb the $v$ zero modes, one can easily see that
their contributions factorized from the correlation function at the
locus of $\dw$ and $\db$ fixed points. Thus, the original topological
interpretation can be maintained. However, the use of consistent
anomaly seems to be restricted to the zero modes of $v$ due to the Gribov
ambiguity\Gri, and they have many other unpleasant aspects\Park.
This motivate us to investigate the covariant forms for
the consistent anomalies.

{}From the density of the covariant anomalies in Yang-Mills theory\BZ,
one can readily obtain the density of covariant anomaly
counterpart in TYM as
\eqn\toccata{
(n+1)c_{n+1} \tr(v\CF^n) \rightarrow (n+1)c_{n+1}\tr(v\hCF^n)
\equiv \uCW_{2n}{}^1,
}
where the superscript denote $\db$ ghost number. For $n=2$,
the covariant anomaly counterpart $\uCW_{4}{}^1$
which can be written in the components
\eqn\messiaen{\eqalign{
{\uCW}_{4}{}^{1,0}&=\ha \tr\left(v\,\CF^2\right),\cr
{\uCW}_{3}{}^{1,1}&=\ha \tr\left(v(\CF\,\Psi+\Psi\,\CF)\right),\cr
{\uCW}_{2}{}^{1,2}&=\ha
                         \tr\left(v(\CF\,\Phi+\Phi\,\CF+\Psi^2)\right),\cr
{\uCW}_{1}{}^{1,3}&=\ha \tr\left(v(\Psi\,\Phi+\Phi\,\Psi)\right),\cr
{\uCW}_{0}{}^{1,4}&=\ha \tr\left(v\,\Phi^2\right).\cr
}}
Then a covariant anomaly is given by
\eqn\gigue{
\uW^{1,4-k}\equiv \int_{\g_{{}_k}}{\uCW}_{k}{}^{1,4-k},
}
where $\g_{{}_k}$ is a $k$ dimensional homology cycle of $M$.

Note that a covariant anomaly, which is gauge invariant,
is not closed under the action
of $\db$ operator ($\db v=-v^2)$
\eqn\pig{
\db\uW^{1,2n-k}(v,\cdot) = - \uW^{2,2n-k}(v^2,\cdot).
}
The above transformation rule motivate us to redefine the $\db$ algebra
slightly such that
\eqn\monkey{\eqalign{
&\tb \CA = -dv -\{\CA,v\},\cr
&\tb v =0,\cr
&\tb\bar v =\pi,\cr
&\tb\pi =0,\cr
&\tb \hCF =-[v,\hCF].\cr
}}
Note that the replacement $\db\rightarrow\tb$ does not change the fixed
point of BRS symmetry $\dA v=0$ and $\tb$ is nilpotent at the fixed point
$\tb^2 \CA = \dA v^2$.
The $\tb$ operator can be regarded a covariant
derivative for $\db$ acting only on the basic  $\db$ triplet
$(v,\bar{v},\pi)$.
That is,
if we introduce the anti-ghost $\bar v$ and auxiliary field $\pi$
and use a particular $\db$ algebra as
\eqn\canon{
\db v = -v^2,\quad \db \bar{v} = \pi -[v,\bar{v}],\quad
\db\pi= -[v,\pi],
}
we can read off corresponding $\tb$ algebra as \monkey. Thus,
deformation of $\db$ to $\tb$ change essentially nothing.
Now the covariant anomaly is $\tb$-closed.

Though  $\uW^{1,4-k}$ are not $\dw$ closed in
general, one can easily see that $\uW^{1,4}=3c_{3}
\tr(v\, \Phi^2)$ is $\dw$ closed and non-trivial $(\dw v=0)$.
The situation is quite similar to the Witten's basic observable
$\tW^{0,4}=c_2\tr \Phi^2$ which is gauge invariant, $\tb$ and
$\dw$ closed and non-trivial. This motivate us to find
an analogue of the topological descent equation \fuga.
One can easily find that
\eqn\haiena{
d \tr(v\,\Phi^2) = - \dw \tr(v(\Psi\Phi+\Phi\Psi)) +\tr (\dA v\,\Phi^2).
}
Repeating the same procedure as \fuga, one can derive a descent equation
\eqn\eagle{\eqalign{
                0&= -\dw\uCW_{0}{}^{1,4},\cr
d\uCW_{0}{}^{1,4}&= -\dw\uCW_{1}{}^{1,3}
+ \tr\left(\dA v\,\Phi^2\right),\cr
d\uCW_{1}{}^{1,3}&= -\dw\uCW_{2}{}^{1,2}
+ \tr\left(\dA v(\Psi\,\Phi+\Phi\,\Psi)\right),\cr
d\uCW_{2}{}^{1,2}&= -\dw\uCW_{3}{}^{1,1}
+ \tr\left(\dA v(\CF\,\Phi+\Phi\,\CF+\Psi^2)\right),\cr
d\uCW_{3}{}^{1,1}&= -\dw\uCW_{4}{}^{1,0}
+ \tr\left(\dA v(\CF\,\Psi+\Psi\,\CF)\right),\cr
d\uCW_{4}{}^{1,0}&=
  \tr\left(\dA v\,\CF^2\right).\cr
}}

The above descent equation contains an extra term unlike
the topological descent equation\fuga, which
vanishes at the $\tb$ fixed point  $\db\CA=\tb\CA = 0$.
After integrating the $i$th relation above over a $i-1$ dimensional
cycle $\g$,   we can see that
$\uW^{1,4-k}$ is $\dw$ closed
\eqn\horse{
\dw\uW^{1,4-k} = 0
}
at the $\db$ fixed point. If we integrate the extra term by part,
we can see that the $\uW^{1,4-k}$ is $\dw$ closed at
the $\dw$ fixed points $\dw\Psi=\dw\chi=0$.
Then, we can conclude
that a covariant anomaly $\uW^{1,4-k}$ is $\dw$ closed
if we restrict the configuration space the locus of $\dw$ and $\db$
fixed points.
One can also easily find that $\uW^{1,4-k}$ is
non-trivial and its $\dw$ cohomology class depends only on the
homology class of $\g_{{}_k}$ as the Witten's observables.
Thus, the covariant anomaly can be used as an topological observable as
long as the reduction of the theory to the locus of BRST fixed points
is exact.
One can also start from $\uCW_{0}{}^{1,2n} =(n+1)c_{n+1}\tr (v\,\F^n)$
and obtain $\uCW_{k}{}^{1,2n-k}$,
which satisfy the same kind of descent equation \eagle. Clearly,
an arbitrary $\uW^{1,2n-k}$ is $\dw$ closed at the locus of the BRST
fixed points.

Let $\triangle U$ and $\triangle u$ denote the net violation of
$\dw$ and $\dt$ ghost number, respectively, due to the zero-modes.
Then the correlation function
\eqn\prelude{
\left<
\prod_{i=1}^{r}\tW^{0,4-k_i}
\prod_{j=1}^{\triangle u}\uW^{1,4-\ell_j}\right>
}
will be non-zero for
\eqn\mazuruka{
\triangle U= h^1-h^0=\sum_{i=1}^{r}(4-k_i)
+ \sum_{j=1}^{\triangle u}(4-\ell_j),
}
where we will always assume that $h^2=0$ and restrict $n=2$ for
simplicity.
In the semiclassical limit the correlation function \prelude\ will
be reduced to the integration of the wedge products of certain closed
differential form over the moduli space of the instanton with the space
of $\F$ and $v$ zero-modes.

Now we will discuss possible branches;
{\bf (A)} There is no reducible connection- thus all possible zero-modes
of $v$ is originated from the Gribov ambiguity and the formal dimension
is the actual dimension.
{\bf (B)} There are reducible connections and all the zero-modes of $v$ is
originated from them - thus the number of $v$ zero-modes
equals that of $\F$ zero-modes.

In the branch A, a zero-mode of $v$
can not be defined globally. It will have different value in the
different coordinate patch and it has no relation with the cohomology
structures of the moduli space.
Thus, inserting
$\uW^{1,4-\ell_j}$ should not in principle affect the global
topological meaning of the correlation function. It will be
sufficient to use $\triangle u$ copies of the $\uW^{1,0}$ only to
control the zero-modes of $v$. Then, one can immediately see that
the correlation function \prelude\ factorized
\eqn\largo{
\left<\prod_{i=1}^{r}\tW^{0,4-k_i}\right>_{\dw \chi=0}
\left<\prod^{\triangle u}\uW^{1,0}\right>_{\db\CA=0},
}
where
$$
\hbox{dim}(\CM) =\triangle U= \sum_{i=1}^{r}(4-k_i),
$$
and $<\cdots>_{\dw \chi=0}$ denotes the integration of the wedge product
of $\dw$ cohomology classes over the moduli space of instanton
$\CM$\W\WW.
The cohomology classes can be obtained by replacing $(\CF,\P,\F)$ in
$\tW^{0,4-k}$ by their (instanton value, zero-modes $\p_i$ , $<\!\F\!>$),
respectively, where $<\!\F\!>$ is given by
$$
<\F> =  -\int_M \Fr{1}{\dA\dA^{*}}[\p,*\p].
$$
Then  $\tW^{0,4-k_i}$  reduces to a closed differential $4-k$ form
on the moduli space
$$
f_{4-k_i} = f_{j_{1}\cdots j_{4-k_i}}\p^{j_{1}}\cdots\p^{j_{4-k_i}}.
$$
The cohomology class $f_{4-k_i}$ is Poincar\'{e} dual to a codimension
$4-k_i$ homology class in $\CM$ and the correlation function
reduce to intersection number of the homology classes,
which is an invariant of smooth four-manifold\D.

The additional term $<\cdots>_{\dw \CA=0}$
denotes the integration of the wedge product of the local $\tb$
cohomology classes over the space of $v$ zero-modes, which may
be related to the group cohomology - note that $\uW^{1,0}$ is a
$\tb$ closed one form in the space of gauge group $\CG$ and $\tb$
cohomology is equivalent to $\db$ cohomology in the BRS fixed points.
Thus, the original topological meaning of the correlation function
is maintained in the branch A.

The theory in the branch B is far more subtle and we do not know the
final answer. What we will try to do is just a formal analysis.
Note that in the branch B
both anti-commuting $v$ and commuting $\F$ zero-modes obey the same kind of
equation $\dA\F=\dA v =0$.
Consequently, the numbers of the $v$ and $\F$ zero-modes
are identical (equal to $h^0={\triangle u}$)
and the spaces of the zero-modes are isomorphic.
Thus, the $v$ and $\F$ zero-modes always appear in the
pair and we can assign a certain supersymmetry between
them.
Another important point is that
the products of the observables in the correlation function \prelude\
obeying the selection rule \mazuruka\ is not the top form in the moduli
space $\CM$, which actual dimension is given by
${\triangle U} + h^0$, unlike the branch A.
Note that the inserted observables carry $\db$ ghost number $h^0$.
The above properties motivate us to examine a possibility to transform the
$\db$ ghost number into $\dw$ ghost number, such that we can obtain the
desired top form in $\CM$.

Note that $v$  and $\F$ have $(\dw,\tb)$ ghost number $(0,1)$ and
$(0,2)$ respectively. Then, we redefine a $\tb$ algebra
\eqn\allegro{
\tb v = -\F.
}
Note that by the modified
$\tb$ algebra the Witten's observables are trivialized. One can easily
prove in general
\eqn\stacato{
\tb \uW^{1,2n-k_i} = -\tW^{0,2n + 2 -k_i}.
}
It does not, however, leads that the correlation function vanishes
because the action $\tb$ on $v$ change the total ghost number by one
while the both ghost number should be preserved independently\M.

If we interpret a $v$ zero-mode $\hv_{i}$ as a one-form (a Grassman
variable) on the space $V$ of $v$ zero-modes, which is a space of
$h^0$ Grassman variables, and a $\f$ zero-mode
$\f_{i}$ as a one-form on the space $W$ of $\F$ zero-modes,
the action of the $\tb$ transformation \allegro\
on $\hv_i$ is an exotic supersymmetry
\eqn\allegreto{
\tb \hv_i = -\f_i.
}
Furthermore, we interpret the action of $\tb$ on $\hv_{i}$
as a Grassmann differential;
\eqn\allegroassai{
\tb \hv_{j}\equiv \left\{\Fr{\rd}{\rd \hv_i},\hv_{j} \right\}\f_i.
}

Note that a Witten's observable $\tW^{0,4-k_i}$ corresponds to a $4-k_i$
form in $\CM$ and a zero-form in $V$, and $\uW^{1,4-\ell_j}$  corresponds
to a $4-\ell_j$ form in $\CM$ and a one-form in $V$.
Thus we use the convention that
$f_{(\ell,k)j_{1}\ldots j_\ell i_{1}\ldots i_k }
\hv_{j_1}\cdots\hv_{j_\ell}\p_{i_1}\cdots\p_{i_k} $
denote a $(\ell,k)$ form on $V\times \CM$.

Now the correlation function \prelude\ reduce to the integral
\eqn\piano{
\int_{\CM}\int_{W}\int_{V}
f_{0,4-k_{i_1}}\wedge \cdots \wedge f_{0,4-k_{i_r}}
\wedge f_{1,4-\ell_{j_1}}\cdots\wedge f_{1,4-\ell_{j_{\triangle u}}}.
}
Note that the integration over $V$ is a Berezin integral and the
integrand is a top form on $V$. Then, we can perform the integral
which is identical to the differentiation. From the relation
\stacato, one can infer that the integral \piano\ reduce to
\eqn\pianosimo{
\int_{\CM}\int_{W} f_{0,4-k_{i_1}}\wedge \cdots \wedge f_{0,4-k_{i_r}}
\wedge f_{0,5-\ell_{j_1}}^1\cdots\wedge f_{0,5-\ell_{j_{\triangle
u}}}^1,
}
where the superscript $1$ denotes that  $f_{0,\ell_{5-j_n}}^1$ is a
one-form in $W$.
After integrating over $W$ eq.\ \pianosimo\ reduces to
\eqn\integral{
\int_{\CM} f_{0,4-k_{i_1}}\wedge \cdots \wedge f_{0,4-k_{i_r}}
\wedge \tilde{f}_{0,5-\ell_{j_1}}\cdots\wedge
\tilde{f}_{0,5-\ell_{j_{\triangle u}}}.
}

Now the integrand of the integral \integral\ is a top-form in $\CM$,
however, the problem is to determine whether it is a
topological invariant. Because the space $W$ of $\F$ zero-modes
is non-compact in a serious way, the convergence of the integrals
\pianosimo\integral\ can not be guaranteed, and because the moduli
space $\CM$ has singularities,
the semi-classical approximation becomes doubtful\WW.
Note that we have inserted a new set of observables into the correlation
function because of the singularities. Then, there can be two possibilities;
{\bf (a)} the new set of observables $\uW^{1,2n-k_i}$ in general may acts
as a regulator for the singularity such that we can maintain the topological
meaning of eq.\ \integral,
{\bf (b)} the semi-classical exactness of the path integral is
broken by the singularity such that the new set of observables is
no longer $\dw$ invariant, then eq.\ \integral\ is no longer
topological. If the second possibility is the case, we can conclude that
the singularity of $\CM$ induce a topological symmetry breaking
mechanism. However, I do not know the answer.

Finally, I will just mention that we can generalize the gauge fixed action
$L$ of TYM to a perturbed one
\eqn\coda{
L\,\rightarrow L\, + \sum_{\a} t_\a\,\tW^{0, \a}
+\sum_{\b} s_\b\, \uW^{1, \b},
}
where $t_\a,s_\b$ are some deformation parameter. Then the partition
function for the perturbed action will reduce to the correlation function
of an appropriate set of observables of the unperturbed theory\tg.
\vfill
\eject
\listrefs
\end